\documentclass[pre,twocolumn]{revtex4}
\usepackage{psfrag}
\usepackage{amssymb,amsmath,amsthm}
\usepackage[dvips]{graphicx}
\usepackage{verbatim}
\newcommand{\beq}{\begin{equation}}
\newcommand{\eeq}{\end{equation}}
\newcommand{\bea}{\begin{eqnarray}}
\newcommand{\eea}{\end{eqnarray}}
\begin{document}
\title{MEASURING DISLOCATION DENSITY IN ALUMINUM WITH RESONANT ULTRASOUND SPECTROSCOPY}
\author{FELIPE BARRA$^{1,2}$, ANDRES CARU$^{1,2}$, MARIA TERESA CERDA$^{1}$, RODRIGO ESPINOZA$^{2}$\footnote{Current address: EMAT, Universiteit Antwerpen, Groenenborgerlaan 171 B-2020, Antwerp, Belgium}, ALEJANDRO JARA$^{1}$, FERNANDO LUND$^{1,2}$ and NICOLAS MUJICA$^{1,2}$}
\affiliation{$^1$DFI and $^2$CIMAT, Facultad de Ciencias F\'\i sicas y Matem\'aticas, \\ Universidad de Chile,
Av. Blanco Encalada 2008, Santiago, Chile.}

\begin{abstract}

\noindent Dislocations in a material will, when present in enough numbers, change the speed of propagation of elastic waves. Consequently, two material samples, differing only in dislocation density, will have different elastic constants, a quantity that can be measured using Resonant Ultrasound Spectroscopy. Measurements of this effect on aluminum samples are reported. They compare well with the predictions of the theory.

\bigskip


\end{abstract}

 \maketitle


\section{Introduction}
\label{intro}
A series of papers by Maurel {\it et al.} [2004a; 2004b; 2005a; 2005b; 2006; 2007a; 2007b] have constructed a detailed theory of the interaction of elastic waves with dislocations in elastic, homogeneous and isotropic, solids. This has been done both in two and three dimensions. Dislocations have ben considered both in isolation as well as in large numbers. In the latter case,  generalization of the Granato \& L\"ucke [1956a; 1956b; 1966] theory has emerged with results for change in wave propagation velocity and attenuation length that clearly distinguish between longitudinal (acoustic) and transverse (shear) polarizations. These results are in satisfactory agreement with laboratory measurements of acoustic attenuation using Resonant Ultrasound Spectroscopy---RUS [Ledbetter \& Fortunko, 1995; Ogi {\it et al.,} 1999; 2004; Ohtani {\it et al.,} 2005]. The case of an isolated dislocation in a half-space has also been studied, as well as the case of low angle grain boundaries mimicked as dislocation arrays.

\vspace{0.5em}
A natural development of the above ideas and results is to ask whether RUS can be turned into a practical tool to measure dislocation densities in materials. In order to do this, it is necessary to validate whatever results are obtained using RUS with time-tested, but more involved, High Resolution Transmission Electron Microscopy [Williams \& Carter, 2004; Arakawa {\it et al.}, 2006; Robertson {\it et al.}, 2008]. A preliminary step towards that aim is to perform RUS measurements on a number of  samples of a given material, one as received from the provider, and the others after cold rolling, or annealing, so as to have significantly different dislocation densities. This paper reports a simplified form of the theory, as well as the first such measurements, using aluminum.

\section{Effective velocity of elastic waves in a dislocation-filled medium}
\label{sec_theo}
We study an homogeneous, isotropic, three-dimensional, infinite elastic medium of density $\rho$, whose state is described by a vector field $\vec u (\vec x, t)$, the displacement $\vec u$ at time $t$ of a point whose equilibrium position is at $\vec x$. In the absence of dislocations, the displacement $\vec u$ obeys the wave equation
\beq
\rho \frac{\partial^2 u_i}{\partial t^2}  -
c_{ijkl}\frac{\partial^2 u_k}{\partial x_j \partial x_l} = 0
\label{eq1}
\eeq
with $c_{ijkl} = \lambda \delta_{ij} \delta_{kl} + \mu (\delta_{ik} \delta_{jl} + \delta_{il} \delta_{jk})$ the tensor of elastic constants, and $i,j,k, \dots = 1,2,3$. A consequence of this equation is that the medium allows for the propagation of longitudinal (acoustic) and transverse (shear) waves with propagation velocity  $c_L = \sqrt{(\lambda +2\mu)/\rho}$ and $c_T = \sqrt{\mu/\rho}$, respectively. Their ratio $\gamma = c_L/c_T > 1$ is always greater than one.

\vspace{0.5em}
Dislocations are modeled as one dimensional objects (``strings'', [Koehler, 1952; Granato \& L\"ucke, 1956a,b]) $\vec X (s,t)$, where $s$ is a Lagrangean parameter that labels points along the line, and $t$ is time, of length $L$, pinned at the ends, whose equilibrium position is a straight line. They are characterized by a Burgers vector $\vec b$, perpendicular to the equilibrium line. Their unforced motion is described by a conventional vibrating string equation
\beq
m\frac{\partial^2 X_i}{\partial t^2} + B \frac{\partial X_i}{\partial t}
-\Gamma \frac{\partial^2 X_i}{\partial s^2} = 0
\label{eq2}
\eeq
where the mass per unit length $m$ and line tension $\Gamma$ are given by [Lund, 1988]
\bea
m & = & \frac{\rho b^2}{4\pi} \left( 1+\gamma^{-4} \right) \ln \left( \frac{\delta}{\delta_0} \right) \nonumber \\
\Gamma & = & \frac{\rho b^2c_T^2}{4\pi} \left( 1-\gamma^{-2} \right) \ln \left( \frac{\delta}{\delta_0} \right)
\label{eqml}
\eea
where $\delta$ and $\delta_0$ are external and internal cut-offs.
The coefficient $B$ is a phenomenological term that describes the internal losses of the string due to, for example, interactions with phonons and electrons. We shall only consider glide motion, that is, motion parallel to the Burgers vector $\vec b$.

\vspace{0.5em}
When elastic waves and dislocations interact, both Eqns. (\ref{eq1}) and (\ref{eq2}) acquire right hand side---source---terms, whose structure has been discussed in detail by Maurel {\it et al.} [2005a,b]. Elastic waves in the presence of $N$ dislocations are best described
not in terms of particle displacement $\vec u$ but in terms of particle velocity $\vec v = \partial \vec u / \partial t$ and the wave equation (\ref{eq1}) becomes
\beq
\rho \frac{\partial^2 v_i}{\partial t^2}  -
c_{ijkl}\frac{\partial^2 v_k}{\partial x_j \partial x_l} = s_i
\label{eq1s}
\eeq
where the source term $s_i$ is given by
\bea
 s_i(\vec x,t) & = & c_{ijkl}\epsilon_{mnk} \sum_{n=1}^N\int_{\cal L}
d s \; \dot X_m^n( s ,t) \tau_n b_l  \nonumber \\
 & & \mbox{} \times \frac{\partial}{\partial x_j}
\delta ( \vec x-\vec X^n (s,t) ).
\label{termesourceN}
\eea
where $\epsilon_{mnk}$ is the completely antisymmetric tensor of order three, and $\hat{\tau}$ is a unit tangent along the dislocation line. The string equation (\ref{eq2}) is written for the component of motion along the glide direction, $X \equiv X_k t_k$ and, loaded by a Peach-Koehler [1950] force it becomes
\begin{equation}
m \ddot{X}( s ,t)+ B\dot{X}( s ,t)- \Gamma X''( s ,t)= \mu b \; {\mathsf
  M}_{lk}\partial_l u_k( \vec X ,t) ,
\label{eqmouv2}
\end{equation}
with ${\mathsf M}_{lk}\equiv t_l n_k+t_kn_l$ and $\hat n \equiv \hat{\tau} \times \hat t$ a unit binormal vector. Overdots mean time derivatives, and primes mean derivatives with respect to $s$.

\vspace{0.5em}
At this point it becomes profitable to go to the frequency domain. The loaded string equation (\ref{eqmouv2}) can be solved in terms of normal modes, and the solution plugged into the right hand side of the wave equation (\ref{eq1s}). In the long wavelength limit, $\lambda \gg L$, and for small string displacements, the result of this  operation is
\beq
-\rho \omega^2 v_i   -
c_{ijkl}\frac{\partial^2 v_k}{\partial x_j \partial x_l} = V_{ik} v_k
\label{eqmany}
\eeq
where
\begin{equation}
V_{ik}=  \frac{8 L}{\pi^2}\frac{(\mu b)^2}{m}
\frac{S(\omega)}{\omega^2}  \; \sum_{n=1}^N {\mathsf M}^n_{ij}
 \frac{\partial}{\partial x_j}  \delta (\vec x- \vec X^n_0 ) \;
{\mathsf  M}^n_{lk}{\frac{\partial}{\partial x_l}}
\label{potentialik}
\end{equation}
and
\begin{equation}
S(\omega) \simeq
  \frac{\omega^2}{\omega^2-\omega_1^2 + i\omega B/m}.
\label{defS}
\end{equation}
with
\[
\omega_1 = \frac{\pi}{L} \sqrt{\frac{\Gamma}{m}}
\]
the frequency of the fundamental mode of the string with fixed ends.

\vspace{0.5em}
Maurel {\it et al.} [2005b] have provided two derivations of effective velocities for elastic waves described by Eqn. (\ref{potentialik}). Here we give a third, with a reasoning similar to the one used to study waves in plasmas [Stix, 1992]: The right-hand-side (\ref{potentialik}) is smoothed through the replacement of the discrete sum over dislocation segments by an integral over space with a continuous density $n(\vec x)$ of dislocation segments, and the tensor ${\mathsf M}^n_{ij} {\mathsf  M}^n_{lk}$ by its angular average, assuming all directions equally likely. The last operation can be found in Appendix C of Maurel {\it et al.} [2005b]. Eqn. (\ref{eqmany}) thus becomes, in the case of uniform dislocation density $n(\vec x) = n$,
\bea
\lefteqn{-\rho \omega^2 v_i  -
c_{ijkl}\frac{\partial^2 v_k}{\partial x_j \partial x_l} =}  \nonumber \\
 & & \frac{8}{\pi^2}\frac{(\mu b)^2}{m}
\frac{S(\omega)}{\omega^2} \frac{nL}{15} (3 \delta_{ik} \delta_{lj} + \delta_{il} \delta_{kj}) \frac{\partial^2 v_k}{\partial x_j \partial x_l}
\eea
In wave number space this is an equation
\beq
{\cal O}_{ik} v_k = 0
\label{eigenfreq}
\eeq
with
\begin{eqnarray*}
{\cal O}_{ik} & \equiv & -\rho \omega^2  \delta_{ik} +
c_{ijkl} k_j k_l  + \frac{8}{\pi^2}\frac{(\mu b)^2}{m}
\frac{S(\omega)}{\omega^2} \frac{nL}{15}  \\
 & & \mbox{} \times (3 \delta_{ik} \delta_{lj} + \delta_{il} \delta_{kj}) k_j k_l .
\end{eqnarray*}
The dispersion relation for the elastic waves in this averaged medium is given by the vanishing of the determinant of $\cal O$:
\beq
\det {\cal O} = 0
\label{disprel}
\eeq
For frequencies smaller than the fundamental frequency of the string, $\omega \ll \omega_1$, and small damping, $(\omega B/m) \ll \omega_1^2$, this leads to the following effective longitudinal $(v_L)$ and transverse $(v_T)$ phase velocities:
\bea
v_L & = & c_L \left(
1 - \frac{16}{15\pi^4} \frac{1}{\gamma^2}
\frac{\mu b^2}{\Gamma} \; n L^3 \right) \\
v_T &  = &  c_T
\left( 1 - \frac{4}{5\pi^4}
  \frac{\mu b^2}{\Gamma} \; n L^3
     \right) .
\label{vitesse2}
\eea

\section{RUS measurements}
\label{rus}

\begin{table*}[t!]
\begin{tabular}{|c|c|c|c|c|c|} \hline \hline
Parameter &  Sample 1  &Sample 2 &Sample 3  & Sample 4  &Sample 5  \\ \hline
 Preparation & Annealed $400^\circ$C/10 hrs & Annealed $400^\circ$C/5 hrs & Original  & Rolled at 33\%  & Rolled at 43\%  \\ \hline
$d_1$ [cm]  &$1.701 \pm 0.001$&$1.700 \pm 0.001$& $1.7004 \pm 0.0003$  & $1.696 \pm 0.001$  & $1.7004 \pm 0.0005$ \\
$d_2$ [cm]  & $1.0010 \pm 0.0004$& $0.9992 \pm 0.0004$&  $0.9991 \pm 0.0005$  & $1.000 \pm 0.001$  &  $1.0005 \pm 0.0003$ \\
$d_3$ [cm]  &$5.002 \pm 0.001$ &$5.002 \pm 0.001$ &  $5.000 \pm 0.001$  & $5.000 \pm 0.001$   & $5.001 \pm 0.002$  \\
$\rho$ [g/cm$^3$]& $2.691 \pm 0.002$ & $2.692 \pm 0.002$ &  $2.696 \pm 0.002$ & $2.692 \pm 0.004$ &  $2.691 \pm 0.002$ \\
$C_{11}$ [$10^{11}$ Pa]& $0.868 \pm 0.038 $& $0.880\pm0.023$  &  $0.884 \pm 0.028 $ & $0.888 \pm 0.026$ &  $0.883 \pm 0.023$   \\
$C_{44}$ [$10^{11}$ Pa] &$0.2680\pm 0.0017$ & $0.2664\pm 0.0007$ &  $0.2661 \pm 0.0008 $ &  $0.2647 \pm 0.0007$ &  $0.2643\pm 0.0007$\\
$v_L$ [m/s] &$5679\pm 124$ & $5719\pm 76$ &  $5725 \pm 89 $ &  $5744 \pm 84$ &  $5728\pm 75 $\\  
$v_T$ [m/s] &$3156\pm 10$ & $3146\pm 4$ &  $3142 \pm 5 $ &  $3136 \pm 5$ &  $3133\pm 4 $\\
\hline \hline
\end{tabular}
\caption{Dimensions, density, $C_{11}$, $C_{44}$, $v_L$ and $v_T$ for the five samples. Columns are ordered for increasing expected density dislocation. Absolute errors for $C_{11}$ and $C_{44}$ are computed as the standard deviations of a set of $135=5\cdot3^3$ results: for each of the $5$ positions per sample, the elastic constant seed values used in RUS are varied $3$ times each, within an interval close to the final measured values ($\pm 20\%$). The density is also varied $3$ times within $\langle \rho \rangle \pm \delta\rho$, giving a total of $27$ possible combinations. }
\label{tab:disipacion}
\end{table*}

\begin{figure*}[t!]
 \includegraphics[width =8 cm]{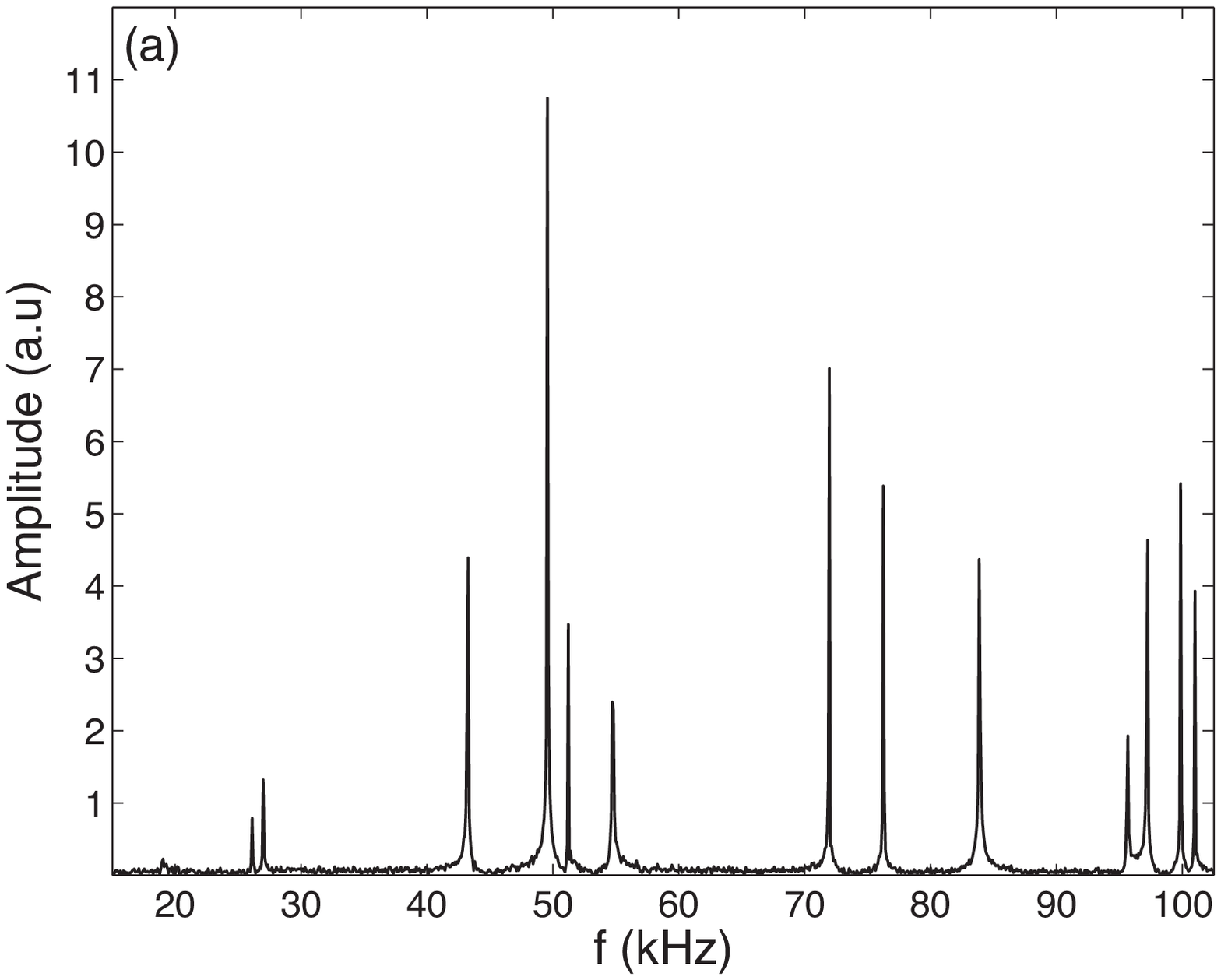}  \hspace{1cm}
 \includegraphics[width = 8 cm]{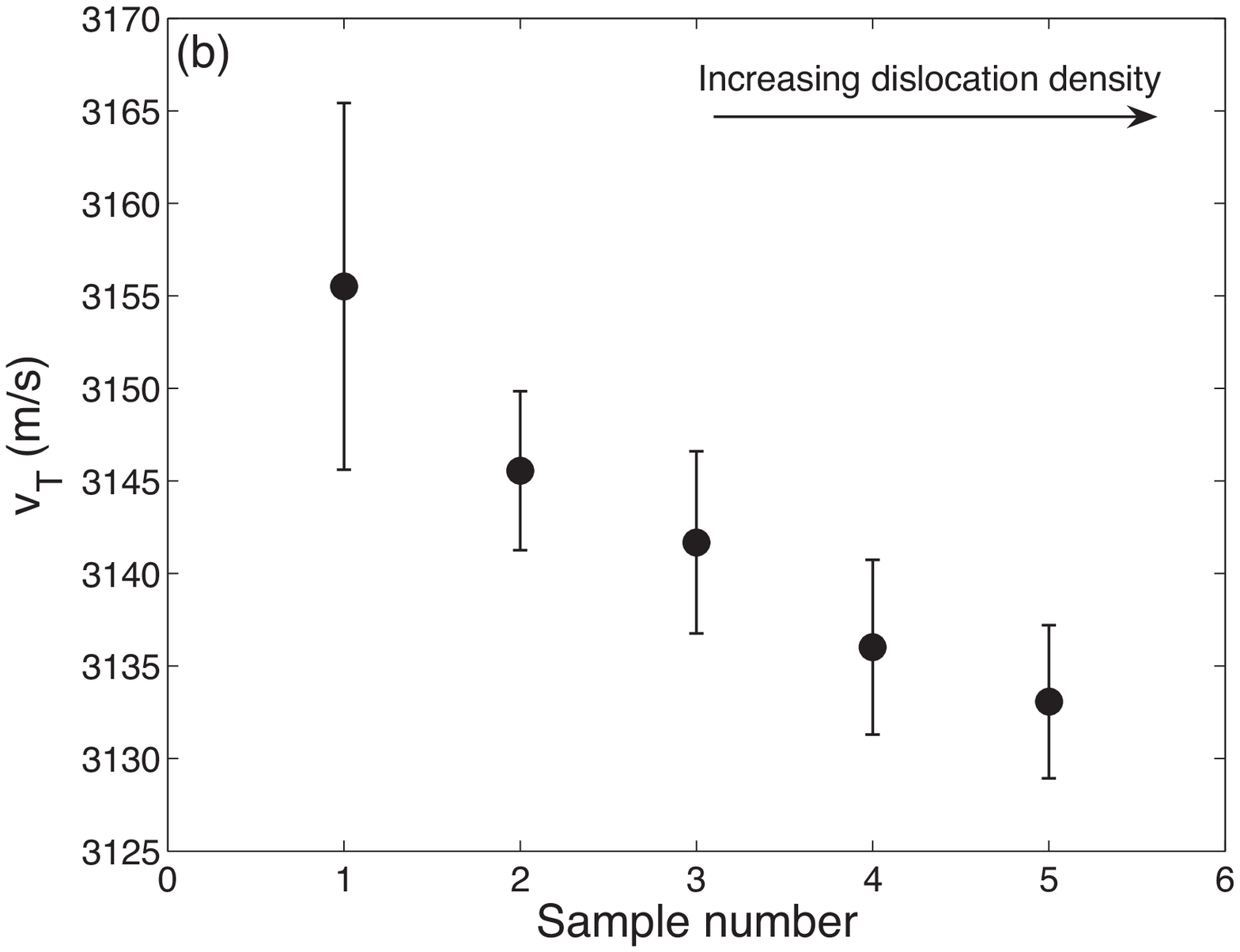}
\caption{(a) Typical frequency spectrum, showing the first 14 resonances for sample N$^\circ$ 3 of Table I. The first resonant mode is at $\approx 18$ kHz, the last one at $\approx 101$ kHz. (b) Shear wave velocity for the five samples under study. The arrow indicates the direction in which dislocation density is expected to increase.}
\label{fig1}
\end{figure*}

Resonant Ultrasound Spectroscopy allows precise measurements of the elastic constants of a sample independent of its symmetries [Migliori {\it et al.}, 1993; Leisure \& Willis 1997]. An homogeneous and isotropic material is characterized by two independent elastic constants, $\lambda$ and $\mu$, or equivalently $C_{11} = \rho v_L^2$ and $C_{44}= \rho v_T^2$.
RUS is known to give more precise measurements of $C_{44}$, and therefore a comparison with the theoretical prediction for $v_T$ is possible.

We have taken five aluminum samples cut from the same bar. One, as bought, one cold-rolled at 33\%, one cold rolled at 43\%, one annealed at 400$^\circ$C, for 5 hours, and another for 10 hours. Longer annealing means lower dislocation density, and stronger cold-rolling means higher dislocation density. The five samples were shaped as parallelepipeds, with dimensions as in Table I. Samples are labeled $1,2,3,4$ and $5$ for increasing expected density dislocation.  A RUS apparatus was built in-house [Car\'u, 2007; Jara, 2007], and used to measure the two elastic constants of the five samples. The sample-apparatus contact force is small, of the order of $0.1$ N. A typical spectrum is shown in Figure 1a. Each resonant frequency was measured for several ultrasonic driving amplitudes (typically five) in order to verify that the resonances are well in the linear acoustic regime. Additionally, each sample was placed five times in the apparatus in order to reduce errors due to slight dependence of the resonant frequencies on the contact load and positioning with respect to the ultrasonic receiver. The measured elastic constants $C_{11}$ and $C_{44}$ as well as the shear and longitudinal wave velocities are also given in Table I. No clear tendency in $v_L$ is observed. Within experimental errors, it is almost constant. However, $v_T$ shows a clear decreasing tendency. It is plotted in Figure 1b versus sample number. The difference between the errors of both wave velocities is consistent with the fact that RUS is much more precise for $C_{44}$, basically because resonant frequencies depend strongly on $C_{44}$ and weakly on $C_{11}$.


\section{Discussion and conclusions}
\label{discussion}
For simplicity we shall assume, a common assumption, $\lambda = \mu$, so that $\gamma^2 =3$. Independent measurements for the original sample give $\gamma^2 = 3.3$, and this ratio does not change significantly from sample to sample.

Using (\ref{eqml}) we have
\[
\frac{\mu b^2}{\Gamma} =  \frac{4\pi \gamma^2}{\gamma^2 -1}  \ln^{-1} \left( \frac{\delta}{\delta_0} \right)
\]
so there is only a dependence on the ratio of cut-offs. Taking $\delta = 10^6 \delta_0$
we get the following expression for the fractional change in shear wave velocity:
\beq
\frac{v_T - c_T}{c_T} = - \frac{16}{5\pi^3} \frac{3}{2} \frac{nL^3}{6\ln 10}.
\eeq
The data of Figure 1b are consistent with
\beq
nL^3 \sim 0.3
\eeq
and with a (linear) trend in agreement with theory, namely that the higher the dislocation density, the lower the effective speed of shear waves.

Taking $L \sim 100$ nm as a typical dislocation length would give a variation in dislocation density among the various samples of order $nL \sim 10^9$ mm$^{-2}$, a conclusion that it should be possible to test by direct measurement with High Resolution Transmission Electron Microscopy (HRTEM). Work along this direction is in progress.

\begin{acknowledgements}
We wish to thank R. Palma and A. Sep\'ulveda for useful discussions. This work has been supported by FONDAP grant 11980002, Anillo ACT N$^\circ$ 15.
\end{acknowledgements}

\bigskip

\noindent {\bf References} \smallskip

\smallskip \noindent
Arakawa, K., Hatakana, M., Kuramoto, E.,
Ono, K. \&  H. Mori, H. [2006], ``Changes in the Burgers Vector of Perfect
Dislocation Loops without Contact with the External Dislocations'', {\it Phys.
Rev. Lett.} {\bf 96}, 125506.

\smallskip \noindent
Car\'u, A. [2007], ``Caracterizaci\'on Ac\'ustica de Materiales'', Acoustics Engineering thesis, Universidad Austral de Chile.


\smallskip \noindent
Granato, A. V., \& L\"ucke, K. [1956a], ``Theory of Mechanical damping due to dislocations'',
{\it J. Appl. Phys.} {\bf 27}, 583-593.

\smallskip \noindent
Granato, A. V., \& L\"ucke, K. [1956b], ``Application of dislocation theory to internal friction phenomena at high frequencies'',
{\it J. Appl. Phys.} {\bf 27}, 789-805.

\smallskip \noindent
Granato, A. V., \& L\"ucke, K. [1966], in {\it Physical
  Acoustics}, Vol 4A,
edited by W. P. Mason (Academic).

\smallskip \noindent
Jara, A. [2007], ``Caracterizaci\'on ac\'ustica de diferentes muestras de aluminio'', (unpublished).

\smallskip \noindent
Koehler, J. S. [1952], in {\it Imperfections in nearly Perfect Crystals},
edited by W. Schockley {\it et al.} (Wiley).

\smallskip \noindent
Ledbetter, H. M. \& Fortunko, C. [1995], ``Elastic constants and internal friction of polycrystalline copper'', {\it J. Mater. Res.} {\bf 10}, 1352-1353.

\smallskip \noindent
Leisure, R. G. \& Willis, F. A. [1997], ``Resonant ultrasound spectroscopy'', {\it J. Phys. Condens. Matter} {\bf 9}, 6001-6029.

\smallskip \noindent
Lund, F. [1988], ``Response of a stringlike dislocation loop
  to an external stress'', {\it J. Mat. Res.} {\bf 3}, 280-297.

\noindent \smallskip
Maurel, A.,  Mercier, J.-F., \& Lund, F. [2004a], ``Scattering of an elastic wave
by a single dislocation'', {\it J. Acoust. Soc. Am.} {\bf  115}, 2773-2780.

\smallskip \noindent
Maurel, A., Mercier, J.-F. \& Lund, F. [2004b], ``Elastic wave propagation through a random array of dislocations'', {\it Phys. Rev. B}
  {\bf 70}, 024303.

\smallskip \noindent
Maurel, A., Pagneux, V., Barra, F. \& Lund, F. [2005a]
``Interaction between an elastic wave and a single pinned dislocation''
{\it Phys. Rev. B} {\bf 72}, 174110.

\smallskip \noindent
Maurel, A., Pagneux, V., Barra, F. \& Lund, F. [2005b]
     ``Wave
     propagation through a random array of pinned dislocations: Velocity
     change and attenuation in a generalized Granato and L\"ucke theory'',
     {\it Phys. Rev. B} {\bf 72}, 174111.

\smallskip \noindent
Maurel, A., Pagneux, V., Boyer, D., \& Lund, F. [2006]
       ``Propagation of elastic waves through polycrystals: the
        effects of scattering from dislocation arrays'',
        {\it Proc. R. Soc. Lond. A},  {\bf 462}, 2607-2623.

\smallskip \noindent
Maurel, A., Pagneux, V., Barra, D. \& Lund, F. [2007a],
``Interaction of a Surface Wave with a Dislocation'', {\it Phys. Rev. B} {\bf 75},
224112.

\smallskip \noindent
Maurel, A., Pagneux, V., Barra, D. \& Lund, F. [2007b]
       ``Multiple scattering from assemblies of dislocation walls in three
       dimensions. Application to propagation in polycrystals'',
        {\it J. Acoust. Soc. Am.},  {\bf 121}, 3418-3431.

\smallskip \noindent
Migliori, A., Sarrao, J. L., Visscher, William M., Bell, T. M., Lei, Ming, Fisk, Z. \& Leisure, R. G. [1993]
 ``Resonant ultrasound spectroscopic techniques for measurement of the elastic moduli of solids'',
     {\it Physica B} {\bf 183}, 1-24.

\smallskip \noindent
Ogi, H., Ledbetter, H.M., Kim, S. \& Hirao, M. [1999],
``Contactless mode-selective resonance ultrasound spectroscopy: Electromagnetic acoustic resonance'', {\it J. Acoust. Soc. Am.} {\bf 106}, 660-665.

\smallskip \noindent
Ogi, H., Nakamura, N., Hirao, M. \& Ledbetter, H. [2004], ``Determination of elastic, anelastic, and piezoelectric coefficients of piezoelectric materials from a single specimen by acoustic resonance spectroscopy'',
  {\it Ultrasonics} {\bf 42}, 183-187.

\smallskip \noindent
Ohtani, T., Ogi, H. \& Hirao, M. [2005], ``Acoustic damping characterization and microstructure evolution in nickel-based superalloy during creep'',
   {\it Int. J. Sol. and Struct.} {\bf 42}, 2911-2928.

\smallskip \noindent
Peach, M. O., \& Koehler, J. S. [1950], ``The Forces Exerted on Dislocations and
the Stress Fields Produced by them'', {\it Phys. Rev.}, {\bf 80}, 436-439.

\smallskip \noindent
Robertson, I. A., Ferreira, P. J., Dehm, G., Hull, R. \& Stach, E. A. [2008],
``Visualizing the Behavior of Dislocations---Seeing is Believing'', {\it MRS
Bulletin} {\bf 33}, 122-131.

\smallskip \noindent
Stix, T. H. [1992], {\it Waves in Plasmas}, Springer.

\smallskip \noindent
Williams, D. B., \& Carter, C. B. [2004], {\it Transmission Electron Microscopy: A Textbook for Materials Science} (Springer, 2004).


\end{document}